\documentclass[conference]{IEEEtran}

\usepackage{amsmath,amsfonts}
\usepackage{algorithm}

\usepackage{algpseudocode}

\usepackage{array}
\usepackage[caption=false,font=normalsize,labelfont=sf,textfont=sf]{subfig}
\usepackage{textcomp}
\usepackage{stfloats}
\usepackage{url}
\usepackage{graphicx}
\usepackage{subcaption}
\usepackage{tabularx}
\usepackage{verbatim}
\usepackage{graphicx}
\usepackage{cite}

\begin{document}
%

\title{Privacy-First Crowdsourcing: Blockchain and Local Differential Privacy in Crowdsourced Drone Services}

\author{
    \IEEEauthorblockN{Junaid Akram\IEEEauthorrefmark{1}, Ali Anaissi\IEEEauthorrefmark{1}\IEEEauthorrefmark{2}}
    \IEEEauthorblockA{\IEEEauthorrefmark{1}School of Computer Science, The University of Sydney, Camperdown NSW 2008, Australia\\
    \IEEEauthorrefmark{2}TD School, University of Technology Sydney, Ultimo NSW 2007, Australia\\
    Email: jakr7229@uni.sydney.edu.au, ali.anaissi@sydney.edu.au}
}


%


\maketitle

\begin{abstract}
We introduce a privacy-preserving framework for integrating consumer-grade drones into bushfire management. This system creates a marketplace where bushfire management authorities obtain essential data from drone operators. Key features include local differential privacy to protect data providers and a blockchain-based solution ensuring fair data exchanges and accountability. The framework is validated through a proof-of-concept implementation, demonstrating its scalability and potential for various large-scale data collection scenarios. This approach addresses privacy concerns and compliance with regulations like Australia's Privacy Act 1988, offering a practical solution for enhancing bushfire detection and management through crowdsourced drone services.
\end{abstract}


\section{Introduction}

The advent of consumer-grade drones has significantly impacted environmental management and emergency services, particularly in bushfire management. These drones enable local residents in fire-prone areas to contribute actively to detection and management efforts through crowdsourced drone services \cite{akram2024DroneSSL, 10492460}. The real-time data collected by drones enhances traditional bushfire management systems by providing comprehensive and timely information \cite{akram2022bc, tahir2022automatic, munawar2022civil}. However, this advancement brings privacy and data protection challenges, as drones gather sensitive information that must comply with stringent legal standards like Australia's Privacy Act 1988. Addressing these privacy concerns is crucial to ensuring broad participation from drone operators and the effective use of crowdsourced data.

Our proposed solution is a privacy-preserving framework that integrates local differential privacy and blockchain technology to create a unique marketplace. In this marketplace, bushfire management authorities serve as data consumers, obtaining crucial information from drone operators who act as data providers. These operators play a vital role in collecting key data, significantly contributing to the generation of important statistics for efficient bushfire detection and management \cite{munawar2022drone, 10535995}. Local differential privacy is a cornerstone of our solution, ensuring robust protection of data providers' privacy against all other system entities, including data consumers and the system operator \cite{akram2023chained}. This method introduces controlled noise into the data, protecting privacy while allowing for useful aggregate information extraction\cite{10547221}.

Additionally, our framework incorporates a blockchain-based solution to facilitate fair and transparent data exchanges. By leveraging Ethereum smart contracts, we establish an immutable log of operations, ensuring that all transactions and data exchanges are recorded and cannot be tampered with. This setup guarantees that drone operators are fairly compensated for their contributions and that bushfire management authorities fulfill their financial obligations transparently. The blockchain also enhances accountability by providing an auditable trail of all transactions, which is crucial for resolving disputes and ensuring trust among participants.

The major contributions of our work include the development of a local differential privacy framework for securely handling crowdsourced drone data, ensuring privacy before data is shared with authorities. We also introduce a blockchain solution to establish an immutable record of operations, ensuring transparent and fair data exchanges. Furthermore, we propose a consent mechanism to foster transparency and encourage voluntary participation from drone operators, crucial for maintaining public trust and adhering to ethical standards. Our paper includes a detailed risk analysis addressing privacy and operational challenges, presenting robust mitigation strategies to ensure the system's resilience and reliability. Lastly, our framework has undergone empirical evaluation through a proof-of-concept implementation, demonstrating its effectiveness and adaptability in real-world scenarios.

\section{System Model}

\begin{figure}
    \centering
    \includegraphics[width=\columnwidth]{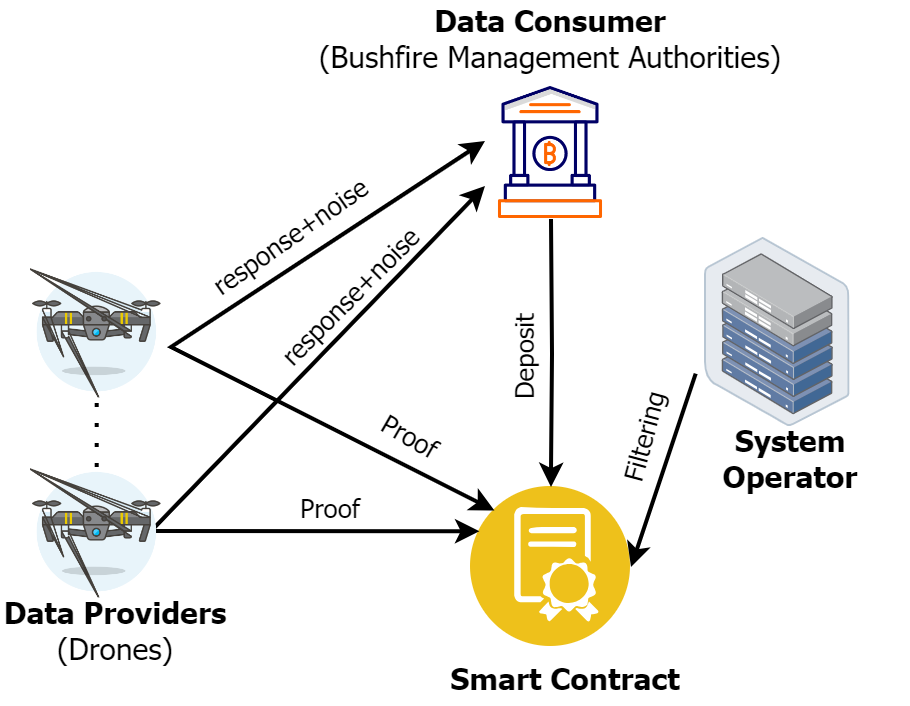}
    \caption{System Overview.}
    \label{fig:figure2}
\end{figure}

Our proposed system, illustrated in Figure \ref{fig:figure2}, comprises key entities tailored for crowdsourced drone data in bushfire management. The bushfire management authority is interested in extracting relevant statistics about fire-prone areas and formulates queries to gather specific data from drone operators. These drone operators are local residents equipped with consumer-grade drones, capable of providing essential data for generating the desired statistics for bushfire detection and management. The system operator is responsible for screening the responses, ensuring drone operators meet the criteria set by the bushfire management authority, and maintaining data accuracy and applicability.

The interaction between entities begins with the bushfire management authority creating a query, conceptualized as a vector \( Q = \{c_1, c_2, \ldots, c_n\} \), representing potential data points relevant to bushfire management. Criteria specifying which drone operators are qualified to respond are attached to this query. Drone operators then indicate their willingness to respond. The system operator assesses and provides data sources that satisfy the requirements. Authorized drone operators, using local differential privacy, send their responses directly to the bushfire management authorities. Finally, the bushfire management authority obtains the required data, pays the system operator, who then compensates the contributing drone operators.

Our system exhibits several key properties. Privacy protection is ensured through local differential privacy, making the precise response of a drone operator unknown to any party, including the system operator. Payment is conditional on service, meaning the bushfire management authority only makes payment after obtaining the specified number of responses. The secure payment process ensures that authorities cannot access any statistics before fulfilling their payment obligations. These objectives are achieved through a blockchain-based system, particularly an Ethereum smart contract, which serves as an unchangeable record and a fair trade platform.

The system assumes specific trust relationships. The bushfire management authority relies on the integrity of drone operators, and due to privacy protection, it is not feasible to confirm the accuracy of individual responses. Trust in system operators is crucial as authorities depend on them for accurate filtering. Public filtering rules and results allow drone operators to submit complaints or rate system operators, fostering accountability. Both authorities and drone operators trust the system operator for fair compensation distribution, supported by the blockchain-based solution. Reputation systems and dispute resolution tools reinforce this trust.

\begin{figure*}[h]
  \centering
  \begin{tabular}[b]{c}
    \includegraphics[width=.21\textwidth]{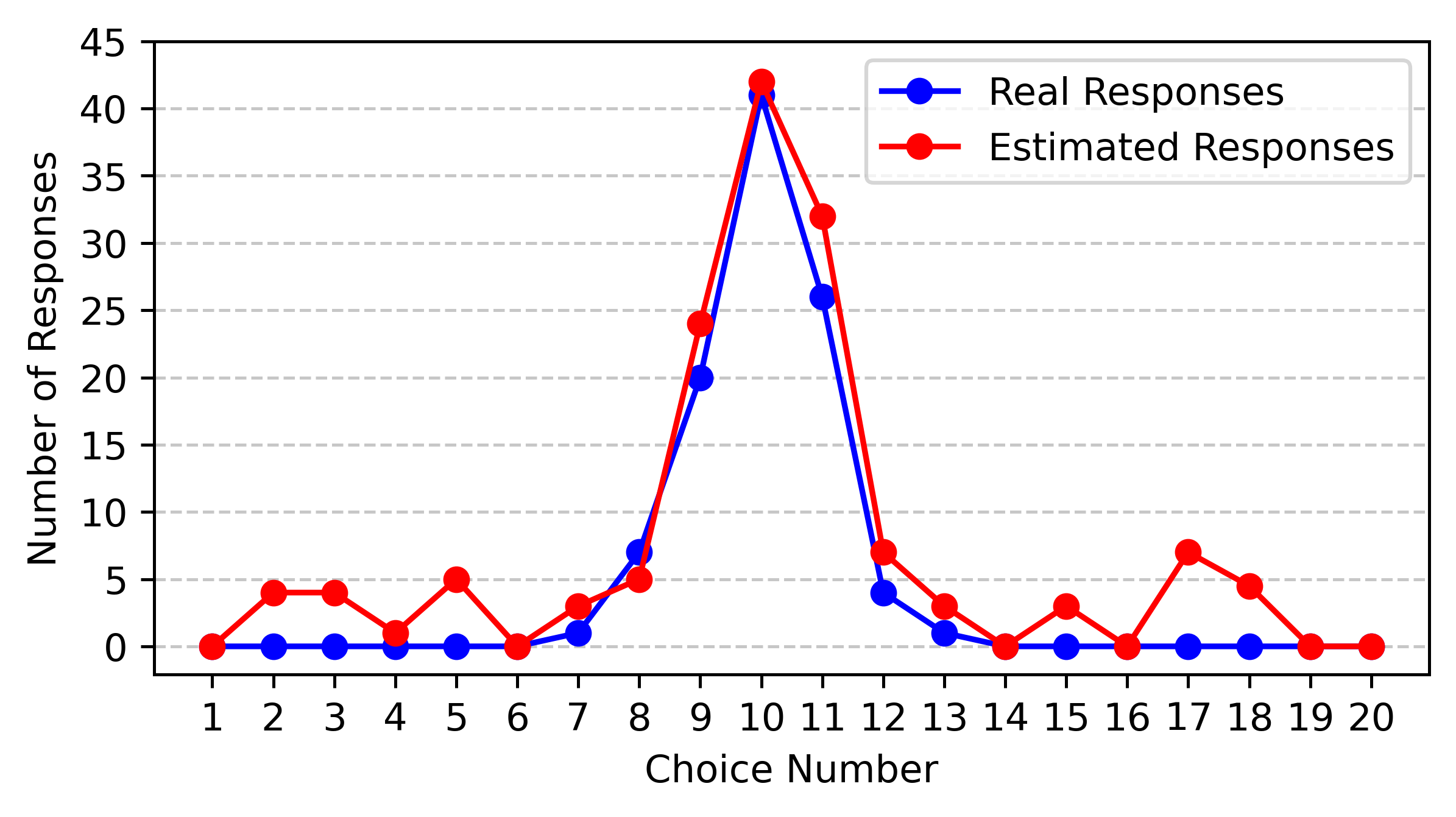} \\
    \small (a) 500 Drones
  \end{tabular} \hfill
  \begin{tabular}[b]{c}
    \includegraphics[width=.21\textwidth]{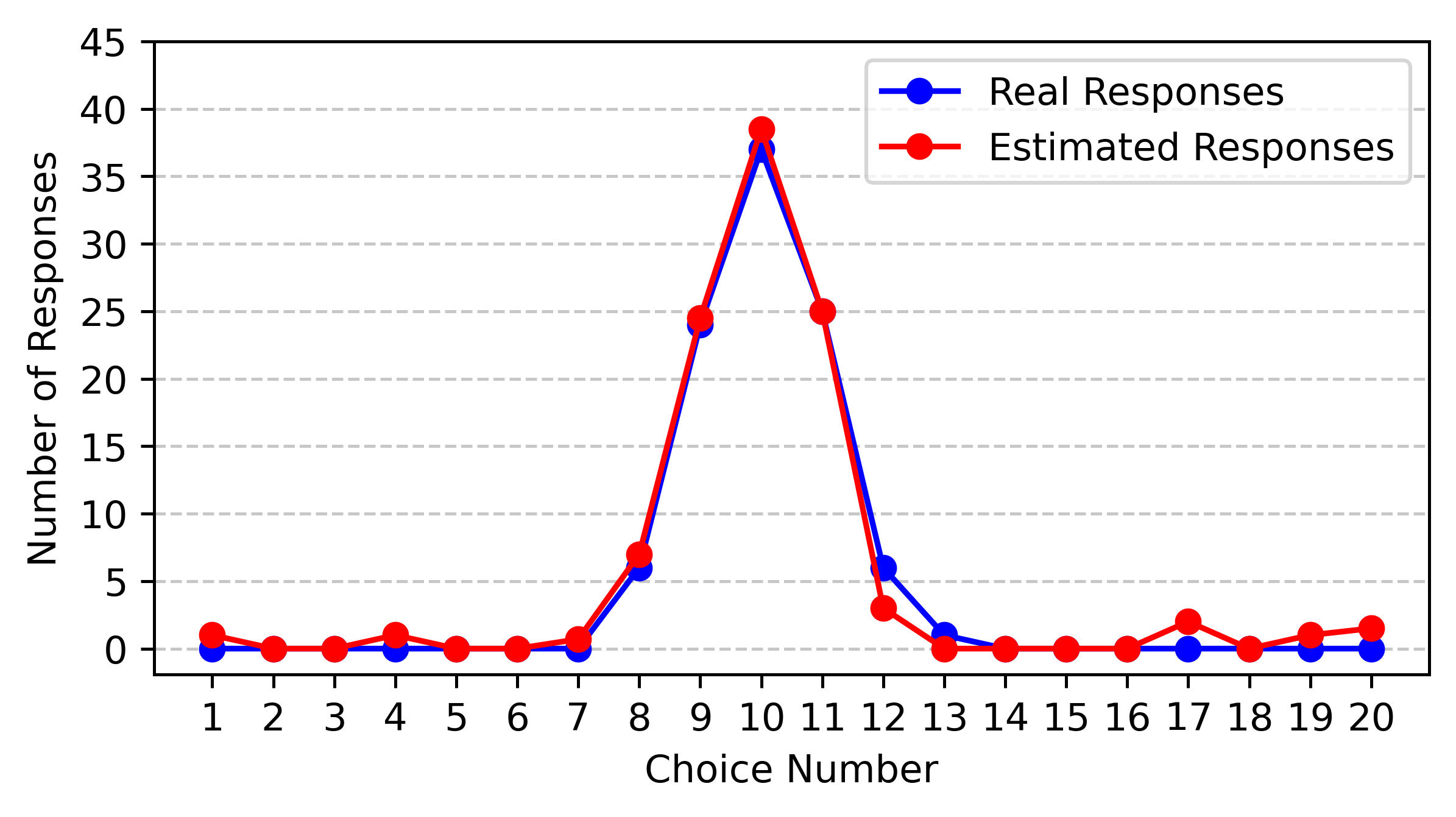} \\
    \small (b) 1000 Drones
  \end{tabular} \hfill
  \begin{tabular}[b]{c}
    \includegraphics[width=.21\textwidth]{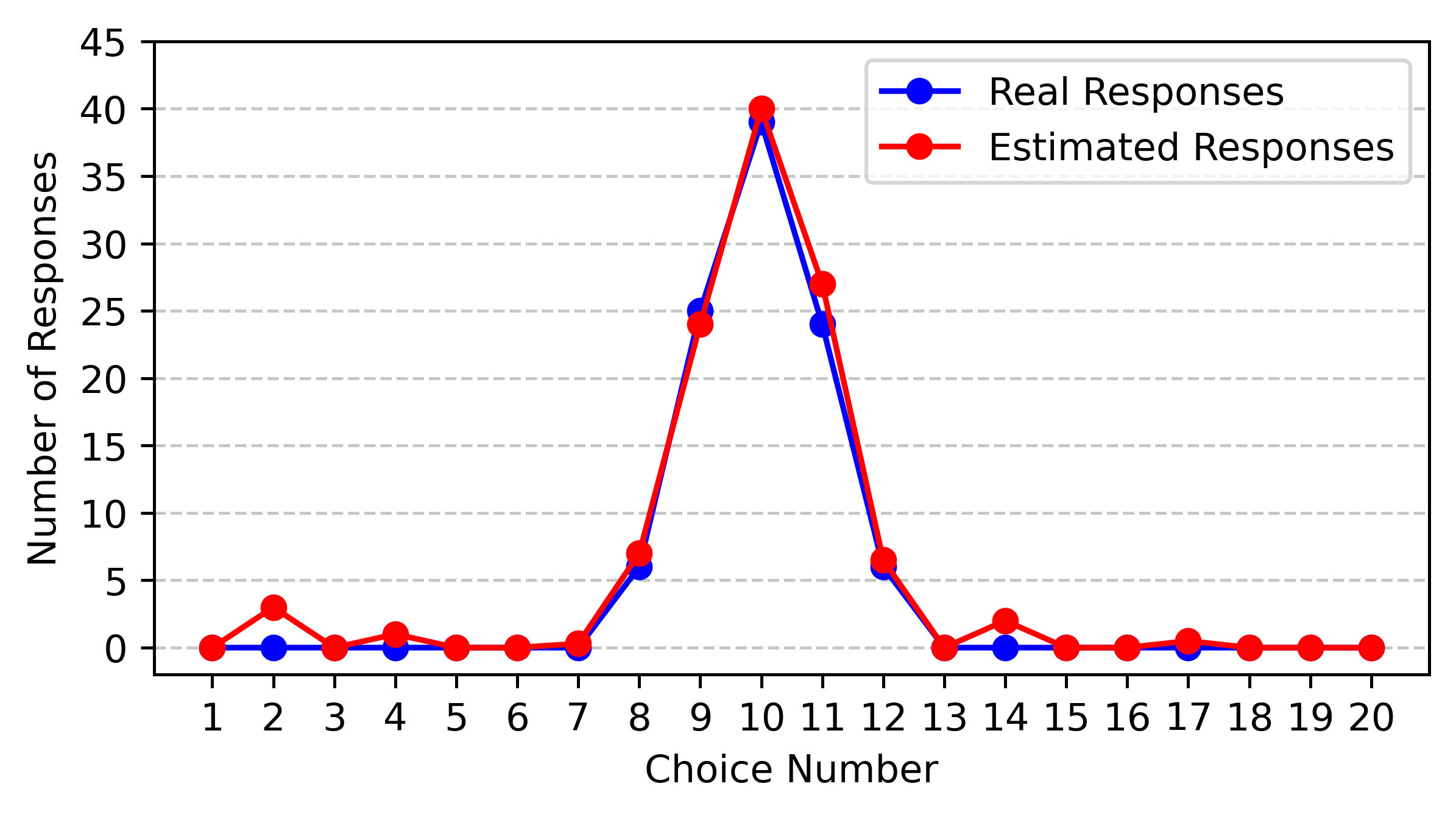} \\
    \small (c) 5000 Drones
  \end{tabular} \hfill
  \begin{tabular}[b]{c}
    \includegraphics[width=.21\textwidth]{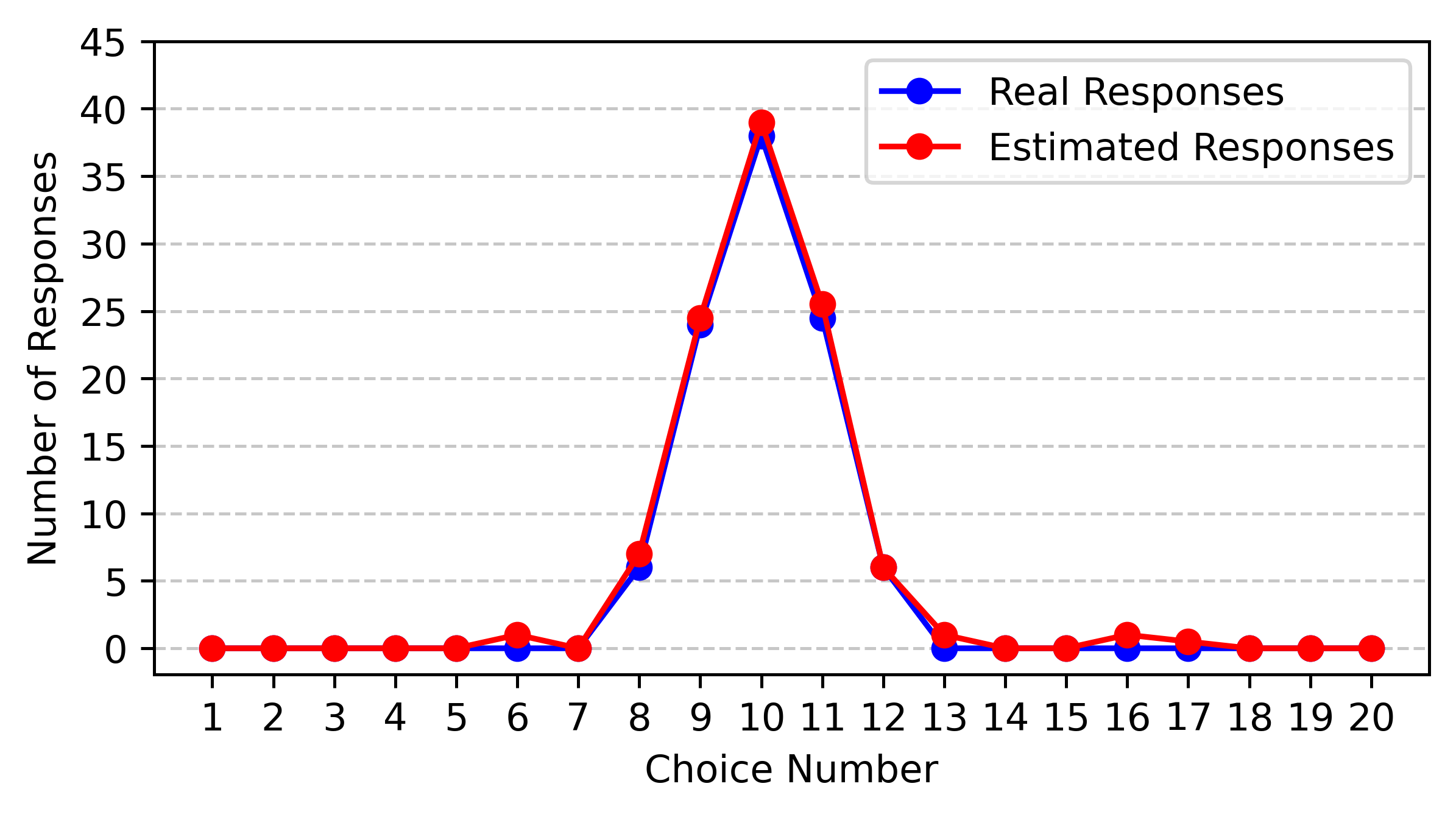} \\
    \small (d) 10000 Drones
  \end{tabular}
  \caption{Impact of local differential privacy on data accuracy with different drone counts. Blue line shows the actual response distribution, red line shows the estimated responses post-privacy implementation.}
  \label{fig:figure3}
\end{figure*}

\section{Design and Implementation} \label{sec_design}

In our system, specifically designed for bushfire management using crowdsourced drone services, we establish a secure communication protocol between the drone operators and the system operator. Each drone operator and the system operator set a pre-shared secret key (psk), a hash function \(H(\text{data})\), and a symmetric encryption method \(E(\text{key}, \text{data})\). Each response is sent from one of the many Ethereum addresses owned by the drone operators, with all current Ethereum addresses known to the system operator.

Bushfire management authorities express their data requirements through specific queries, which may seek information on the location and size of a bushfire. These queries include choices representing different ranges of size or areas. To ensure reliability, the amount and granularity of responses are customized to the requirements of each bushfire management case. A cryptographic hash of a configuration file containing the query and choices is stored on the blockchain, keeping the blockchain costs fixed regardless of the number of choices. Bushfire management authorities can also set specific filtering rules, such as geographical constraints.

The protocol for smart contract creation involves several steps. First, a "survey configuration" is agreed upon by the system operator and the bushfire management authority, comprising the query \( Q = \{c_1, c_2, \ldots, c_n\} \), filtering criteria, and the required number of responses (\( NR \)). A smart contract stores the location and hash of this configuration. There is also consensus on two nonces, \( n_1 \) and \( n_2 \), used to derive an encryption key \( sk = H(s_1 \, || \, s_2) \), where \( s_1 = \text{HMAC}(\text{psk}, n_1) \) and \( s_2 = \text{HMAC}(\text{psk}, n_2) \). The smart contract records the service fee along with \( n_1 \), \( n_2 \), \( H(s_2) \), and \( s_1 \).

Drone operators retrieve the survey configuration, verify its integrity, and prepare their responses using local differential privacy. The fundamental one-time RAPPOR \cite{erlingsson2014rappor} method is employed. Each response \( R \) corresponds to a bit vector representing the query \( Q \). The randomized response game determines the element \( r_i \) in \( R \) for each choice \( c_i \) in \( Q \). The response vector \( R \) is then encrypted with the key \( sk \), creating ciphertext \( C_R = E(sk, R) \), and its hash \( H(C_R) \) is stored in the smart contract. A sorted map \( M_{\text{address} \rightarrow H(C_R)} \) links the drone operator's Ethereum address to its corresponding \( H(C_R) \).

The system operator uses a dynamic filter \( F \) to screen responses. If the associated data source meets the filtering requirements, a bit \( f_i \) in \( F \) is set to 1; otherwise, it is set to 0. Once the number of '1's in \( F \) equals the required number of responses \( N_R \), the system operator finalizes the filtering process, stores the hash \( H(F) \) in the smart contract, and publishes \( F \) at a pre-agreed URL. No more responses are accepted afterward.

After committing their response by entering \( H(C_R) \) in the smart contract, drone operators obtain the filter \( F \) and the map \( M \). If \( f_i = 1 \) for their Ethereum address \( i \), they send the index \( i \) and ciphertext \( C_R \) to the bushfire management authority. The authority verifies \( C_R \) using \( H(C_R) \), confirms eligibility through \( F \), and deposits the agreed amount into the smart contract. The system operator discloses \( s_2 \), and if \( H(s_2) \) matches the stored hash, the contract transfers the deposit to the system operator, who then compensates the drone operators. Finally, the bushfire management authority uses \( s_1 \) and \( s_2 \) to reconstruct \( sk \) and decrypt the responses, deriving significant statistics from the data.

\section{Analysis and Discussion}

The effectiveness of local differential privacy, specifically the RAPPOR algorithm, depends on the number of responses obtained. To evaluate this, we replicated a realistic bushfire data collection scenario by simulating many drone operator instances responding to a query with 20 choices. Each instance's response was selected based on a normal distribution with a mean of 10 and a standard deviation of 2, simulating varied data collected from drones. After submission, we compared the actual distribution of responses against the distribution derived from the noisy responses. Our experiments included trials with 500, 1,000, 5,000, and 10,000 drone operators. The results, shown in Figure \ref{fig:figure3}, indicated that the accuracy of the extracted results remained invariant to the number of choices and the distribution of actual responses, underscoring the robustness of our approach.

Our experiments demonstrated that local differential privacy effectively preserves the utility of the data while protecting individual privacy. Figure \ref{fig:figure3} illustrates the impact of local differential privacy on data accuracy across varying numbers of drone operators. The blue lines represent the actual distribution of responses, while the red lines indicate the estimated distribution based on the noisy responses. The consistency in accuracy, despite increasing the number of responses, highlights the robustness of our approach in accurately capturing and interpreting data for bushfire management purposes.

The results from Figure \ref{fig:figure3} validate the efficiency of our local differential privacy implementation. By maintaining high accuracy in the presence of noise, our system ensures reliable data collection without compromising privacy. This balance between privacy and accuracy is critical for the practical deployment of crowdsourced drone services in bushfire management. Future enhancements may focus on optimizing the algorithm to further improve efficiency and scalability, ensuring the system remains robust and adaptable in various real-world scenarios.

\section{Conclusion}

In this study, we introduced a privacy-preserving marketplace using crowdsourced drone services for bushfire management. Our approach allows authorities to purchase noisy data from drone operators, essential for meaningful bushfire detection statistics. Implementing local differential privacy, we ensure operators' privacy against all entities, including system intermediaries. Our system employs a blockchain-based fair trade mechanism, maintaining an immutable record while minimizing overhead. This design effectively balances privacy, accuracy, and scalability, demonstrating robust performance across various scenarios, ensuring reliable and private data collection for bushfire management.



%

\bibliographystyle{IEEEtran}
\bibliography{ref}

\end{document}